\documentstyle[psfig,12pt]{article}
\title{\bf Inhomogeneous Chemical Evolution\\ 
of the Galactic Halo}
\author{C.~Travaglio$^{1,2}$, A.~Burkert$^1$ and D.~Galli$^3$
\vspace{1cm}\\
\normalsize $^1$Max-Planck Institute f\"ur Astronomie, Heidelberg, Germany\\
\normalsize $^2$Dipartimento di Astronomia e Scienza dello Spazio, Firenze,
Italy\\
\normalsize $^3$Osservatorio Astrofisico di Arcetri, Firenze, Italy}
\date{\mbox{}}
\begin{document}
\maketitle
\pagestyle{empty}
%
%
\def\bull{\vrule height .9ex width .8ex depth -.1ex}
\makeatletter
\def\ps@plain{\let\@mkboth\gobbletwo
\def\@oddhead{}\def\@oddfoot{\hfil\tiny\bull\quad
``WR Stars in the Framework of Stellar Evolution'';
33$^{\mbox{\rm rd}}$ Li\`ege\ Int.\ Astroph.\ Coll., 1996\quad\bull}%
\def\@evenhead{}\let\@evenfoot\@oddfoot}
\makeatother
%
%
\def\beginrefer{\section*{References}%
\begin{quotation}\mbox{}\par}
\def\refer#1\par{{\setlength{\parindent}{-\leftmargin}\indent#1\par}}
\def\endrefer{\end{quotation}}
%
%

{\noindent\small{\bf Abstract:} 
In this contribution we describe the basic features of a Monte Carlo
model specifically designed to follow the inhomogenous chemical
evolution of the Galactic halo, taking into account the effects of
local enrichment and mixing of the halo gas, and with particular
emphasis on elements like Eu produced by $r$-process nucleosynthesis.
We compare our results with spectroscopic data for the chemical
composition of metal-poor halo stars and globular clusters like M13,
M5, M92, M4, and we infer some constraints on the star formation
history of the halo and the rate and mass spectrum of supernovae during
the first epoch of Galaxy evolution.

%
%

\section{Introduction}

The results of several observational studies suggest the existence of a
considerable scatter of heavy element abundances (over about 2 dex in
[element/Fe]) in a large sample of low-metallicity halo stars (see
e.g.  Gilroy et al.~1988, Ryan \& Norris~1991, Gratton \& Sneden~1994,
McWilliam et al.~1995b, Ryan et al.~1996, and, more recently,
McWilliam~1998, Sneden et al.~1998).  However, the actual nature of
this scatter, and the possible influence of various effects like
different calibration methods or data reduction procedures, is still
controversial.  If the observed scatter is {\em intrinsic}, one is led
to consider a scenario in which the oldest halo stars were formed out of
a gas of spatially inhomogeneous chemical composition.  For instance,
the peculiar abundances determined by Sneden et al.~(1994) and
McWilliam et al.~(1995a) in the star CS 22892-052 (where $r$-process
elements are enhanced over 40 times the solar value) strongly support
this hypothesis.  Moreover, the presence of $r$-process elements in
low-metallicity halo stars is indicative of a prompt enrichment of the
Galaxy by early generations of massive stars, as first suggested by
Truran (1981).

A detailed analysis of the Galactic evolution of heavy elements from
Barium to Europium has been recently performed by Travaglio et
al.~(1999) adopting the standard approach to the chemical evolution of
the Galaxy, where stars are assumed to form from a chemically
homogeneous medium at a continuos rate.  As the authors stressed, this
approach is able to reproduce spatially averaged values of element
abundances over the Galactic age, but a more realistic model for the
chemistry and dynamics of the gas is needed in order to investigate the
earliest phases of halo evolution.

Studies of inhomogeneous enrichment of the Galaxy have been recently
carried out by Raiteri et al.~(1999), Tsujimoto et al.~(1999), and
McWilliam \& Searle (1999).  The work by Raiteri et al.~(1999) is
mostly concerned with the evolution of Ba, followed by means of a
hydrodynamical N-body/SPH code; the work by Tsujimoto et al.~(1999) is
focused on the spread in Eu observed in the oldest halo stars,
explained in the context of a model of supernova-induced star
formation; finally, the work by McWilliam \& Searle (1999) is based on
an original stochastic model for the chemical evolution of the Galaxy
aimed at reproducing the observed Sr abundances.

In this contribution we present the results of a Monte-Carlo code for
the chemical evolution of the Galactic halo, based on the idea of
fragmentation and coalescence between interstellar gas clouds. With
this approach, described below in more detail, we take into account the
effect of mixing between clouds, bursts of star formation in the clouds
and the consequent chemical enrichment of the gas, and the delayed
mixing of supernova ejecta into the interstellar medium (hereafter
ISM). In particular, we present here our results for Eu and for the
age-metallicity relation in the earliest phases of the Galaxy.

\section{A Stochastic Model for the Galactic Halo}

The main characteristics of our chemo-dynamical model are ({\em i}\/) a
realistic treatment of chemical inhomogeneities in the ISM due to
incomplete mixing of stellar ejecta, and ({\em ii}\/) the occurrence of
discrete episodes of star formation localized in time and space. Here
we summarize briefly the parameters of our model, and we discuss the
sensitivity of our results to the values adopted for these parameters.

The idea that interstellar clouds collide and grow by coalescence up to
a critical mass at which they become gravitationally unstable and form
stars was first suggested by Hoyle (1953) and Oort (1954).  In the
present work, we consider the halo composed by discrete gas clouds with
initial mass in the range $10^3$--$10^4$~$M_\odot$ and we follow their
evolution for 1~Gyr, with a timestep of $10^6$~yr, since our main
interest is the early chemical evolution of the halo.  Every $10^7$~yr
a cloud experiences a collision with another cloud, with a probability
depending on the mass of the clouds and on their collision cross
section $\sigma_{ij}\propto (M_i + M_j)^{2/3}$.  In particular,
considering clouds as geometrically similar particles, we define a
collision probability
\begin{equation}
P_{ij}=\frac{M_i M_j\sigma_{ij}}
{M_{\rm max}^2\sigma_{\rm max}},
\end{equation}
normalized to the values of the mass $M_{\rm max}$ and cross section
$\sigma_{\rm max}$ of the most massive cloud at each timestep.  As a
result of the process of coalescence, more massive clouds are produced
at each timestep, and at a critical value of 
$M_{\rm cr}=10^4$~$M_\odot$ the cloud becomes gravitationally
unstable and is allowed to give birth to a cluster of stars.  We assume
that the probability for a star formation burst is a function of the
cloud's age and has the shape of a Gaussian peaked on $t=2\times
10^7$~yr after the formation of the cloud. Following a star burst, a
cloud is fragmented by the energetic processes that accompany star
formation, and breaks up into a distribution of smaller clouds.

The mass spectrum of the stars formed in
the burst is described by a Salpeter initial mass function 
\begin{equation}
\frac{dN}{dm}(m)=Am^{-2.35},
\end{equation}
where $m$ is the mass of the star and $A$ is a normalization constant.
We also assume a star formation efficiency $f=0.03$, that is each 
burst converts 3\% of the mass of the parent cloud $M$ into stars.
If the range of stellar mass extends from 0.1~$M_\odot$ to 120~$M_\odot$, 
then $A$ can be determined by the condition
\begin{equation}
A\int_{0.1}^{120}m\frac{dN}{dm}(m)dm=fM,
\end{equation}   
giving
\begin{equation}
A\simeq 0.17f\left(\frac{M}{M_\odot}\right).
\end{equation} 

\begin{figure}
{\vbox{\psfig{figure=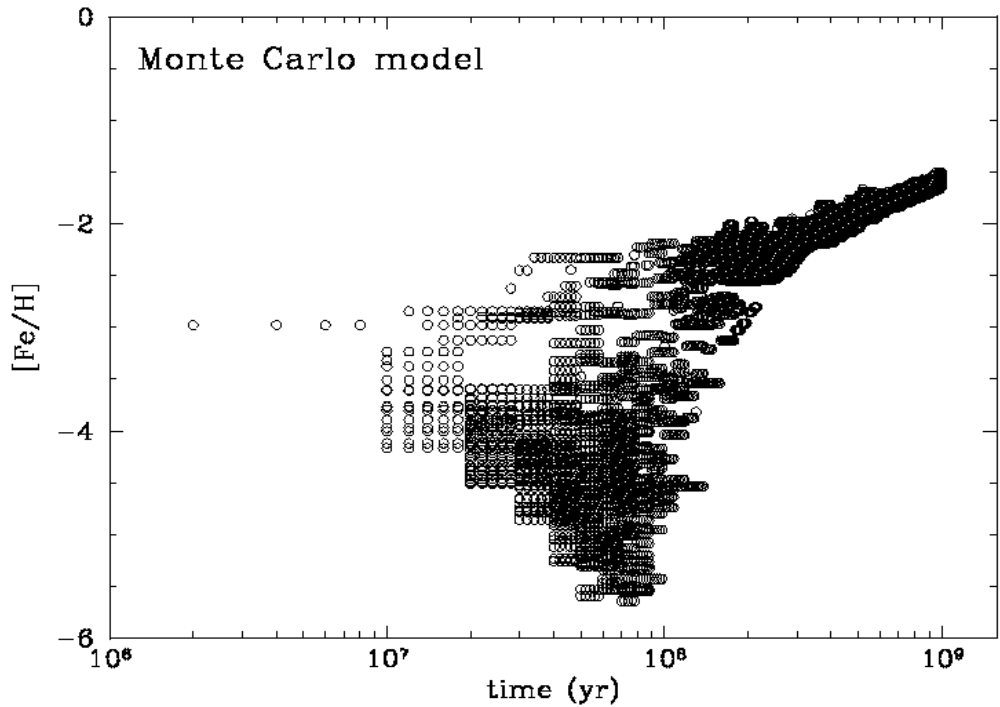,height=10.5cm,width=14.5cm,angle=-90}}}
\caption{Age-metallicity relation obtained with our model. Open circles  
represent interstellar halo clouds during the first Gyr of the
evolution of the Galaxy.}
\end{figure}

As an example of the results of our model, we show in Fig.~1 the
age-metallicity relation obtained under the assumptions described
above. Since Galactic Fe is mostly produced by stars which explode as
core collapse supernovae, we have adopted the Fe yields computed by
Woosley \& Weaver (1995), for the mass range 12--40~$M_\odot$ (model B,
for $Z=10^{-4}Z_\odot$).  As no calculations for progenitors  more
massive than 40~$M_\odot$ are available in Woosley \& Weaver (1995),
for $M > 40$~$M_\odot$ we adopted the yields computed by Woosley,
Langer \& Weaver (1995) for a 60 $M_\odot$ and solar metallicity (model
7K).  We see from Fig.~1 that the first cloud enriched in Fe up to
[Fe/H] $\simeq -3$ appears at $t=2\times 10^6$~yr, following the first
burst of star formation.

With the adopted IMF we can easily derive 
the number ${\cal N}$ of SNe per burst that contribute to this Fe enrichment 
in one cloud, 
\begin{equation}
{\cal N}=\int_{12}^{120}\frac{dN}{dm}(m)dm\simeq 4.2f\left(\frac{M}{10^3~M_\odot}\right).
\end{equation}  
If the cloud's mass $M$ is in the range $10^3$--$10^4$ $M_\odot$
and $f=0.03$, we obtain ${\cal N}\simeq 1$.
This result is in agreement with the point stressed by Ryan, Norris \& 
Bessel (1991) who argued that the ejecta of a single 25 $M_\odot$ 
exploding in a $10^6$~$M_\odot$ cloud is sufficient to enrich a cloud to 
[Fe/H] $= -3.8$, thus setting a lower
limit on the metallicity of the second generation of stars.

From Fig.1 one can also notice that after $10^7$ yr the spread in
[Fe/H] increases, covering a range $-6<$ [Fe/H] $<-2$. After about
$10^8$~yr the halo gas has had enough time to homogenize its chemical
composition, and the spread in [Fe/H] is considerably reduced,
converging to [Fe/H] $\simeq -1.5$.

\section{The Chemical Evolution of Europium}  

Since the pioneering work on stellar nucleosynthesis by Burbidge et
al.~(1957), the origin of nuclei heavier than iron has been attributed
to neutron capture processes, both {\em slow} (the $s$-process), and
{\em rapid} (the $r$-process).  While the $s$-process occurs mainly
during hydrostatic He-burning phases of stellar evolution, the
$r$-process is associated with explosive conditions in SNe.  As first
stressed by Truran~(1981), the abundance of heavy elements in very low
metal-poor stars is compatible with an $r$-process origin. This point
has been recently supported by new observations of low metallicity
stars (see e.g. McWilliam et al.~1995b, McWilliam~1998, Sneden et
al.~1998) and also by the abundance pattern determined in some peculiar
stars like CS 22892-052.

Since Eu is mostly produced by $r$-process nucleosynthesis, the
analysis of the [Eu/Fe] ratio during the early evolution of the Galaxy
can provide constraints both for the inhomogeneous chemical enrichment
of the halo, as well as for the astrophysical site of the $r$-process
elements. The latter, in particular, still needs to be unambigously
identified, despite the large number of recent studies (see e.g. the
hydrodynamic simulations by Wheeler et al.~1998 and Freiburghaus et
al.~1999), and quantitative estimates of the $r$-process yields are
still unavailable.

\begin{figure}
{\vbox{\psfig{figure=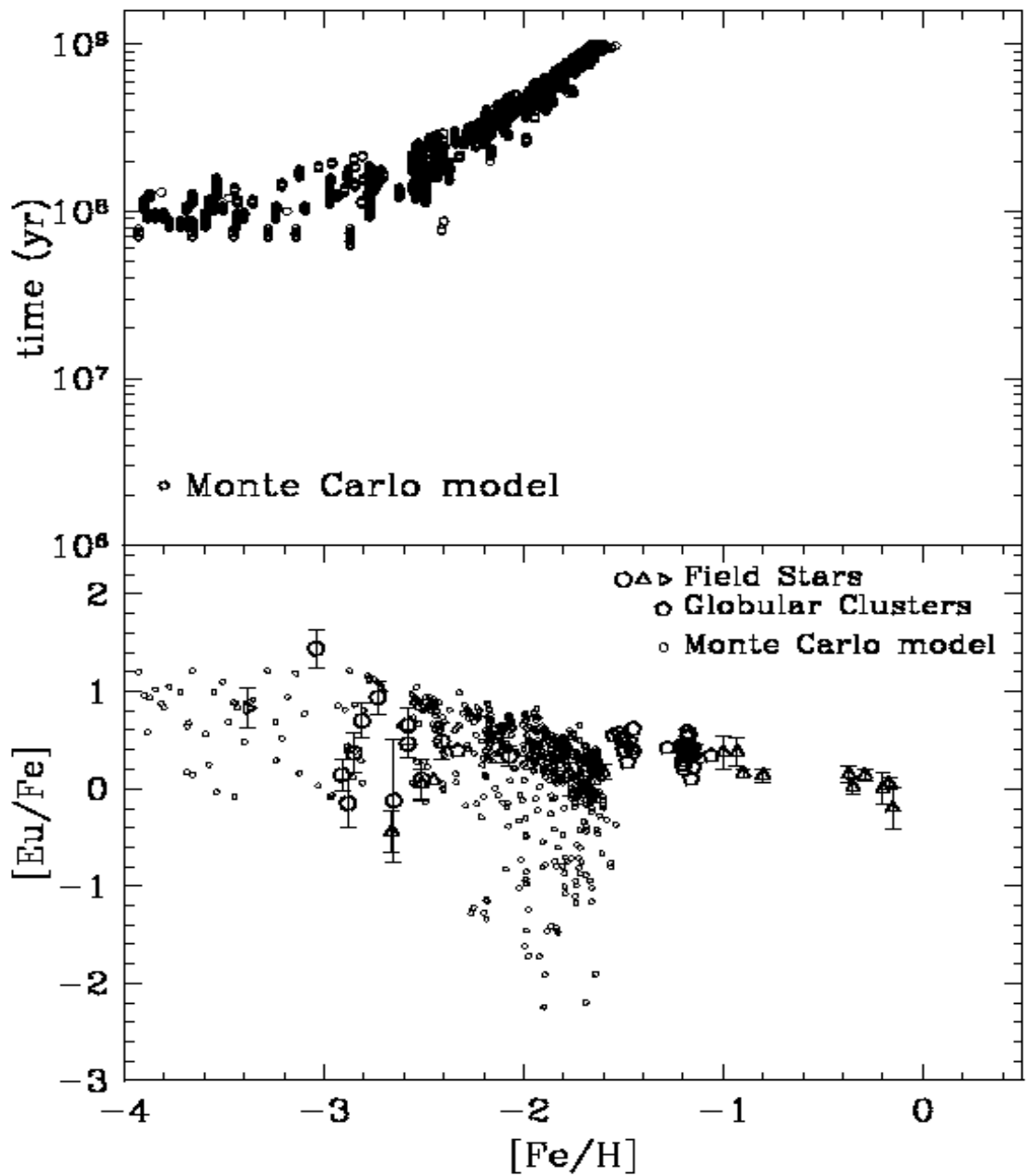,height=19.0cm,width=16.5cm}}}
\caption{{\em Upper panel}: age-metallicity relation as in Fig.~1, 
but only for clouds enriched in Eu. {\em Lower panel}: evolution of Eu
during the first Gyr of the Galaxy, according to our model 
({\it small thin open circles}). 
Observational data of Eu in metal-poor stars are from: 
McWilliam et al.~(1995) ({\em thick open circles}); Gratton \& 
Sneden~(1994)~({\em open triangles});
Norris, Ryan, \& Beers~(1997) ({\em open tilted triangles}). 
Data for Eu in globular clusters stars ({\em open pentagons})
are from Shetrone (1996) for
M13, M5, M92, and from Ivans et al.~(1999) for M4.}
\end{figure}

Recently, the heavy elements enrichment of the Galactic halo has been
analyzed in several studies, e.g. Ikuta \& Arimoto~(1999), Tsujimoto et
al.~(1999), McWilliam \& Searle~(1999). In all these works the
$r$-process yields were deduced empirically from the available
observational data of the most metal-poor stars. In the present work we
adopted the analytical calculation of the $r$-process presented by
Travaglio et al.~(1999), which is independent on observations. These
authors treated the $r$-process as a ``primary'' process originating in
low-mass Type II SNe and derived the $r$-residuals after subtracting
from the solar abundances the predicted $s$-fractions at $t=t_\odot$.

In Fig.~2 (lower panel) we show our result for [Eu/Fe] vs. [Fe/H],
obtained with the model prescriptions described in Sect.~2, and using
the Eu quantitative estimates (from low-mass SNe) by Travaglio et al.
(1999). Our predictions are compared with the available
spectroscopic data for metal-poor field stars.  In the same figure, we
also show, for comparison, the Eu abundances measured in several
globular clusters stars. As stressed by Shetrone (1996), the [Eu/Fe]
ratio follows the same trend with [Fe/H] both in globular cluster stars
and in field stars.  This is a good constraint for the nucleosynthesis
processes that occurred in globular clusters stars. In fact, if both O
and Eu originate from Type II SNe, these elements should be equally
depleted in cluster and field stars, contrary to observational
evidence. Therefore it is likely that the O depletion (observed in
globular cluster stars but not in field stars) is not a consequence of
a primordial effect, but can be attributed to nucleosynthetic processes
during the evolution of these stars.

In Fig.~2 (upper panel) we also show the same age-metallicity relation
shown in Fig.~1, but selecting only clouds in which there are stars
able to produce Eu. This figure can provide more insight about the age
at which the halo gas begins to be enriched by the $r$-process, as well
as the metallicity range that star forming clouds enriched in Eu clouds
can cover at a given epoch.

\section{Discussion}

The results described above concerning the enrichment in Eu of the
Galactic halo make clear that in order to match the considerable spread
in [Eu/Fe] observed in stars in the metallicity range $-3.5 \leq$
[Fe/H] $\leq -2.0$ we need to distinguish the $r$-process sources
(low-mass SNe with $M \simeq 10~M_\odot$) with respect to Fe sources
(in our model Fe is produced by stars in the mass range
12--120~$M_\odot$ during the first Gyr of halo evolution). We also
explored the consequences of zero Fe yields for higher mass stars.
This is supported by the idea that mass-loss increases with metallicity
and that at the lowest metallicities the probability is higher to have
even less Wolf-Rayet stars.  However, since we assume that the chemical
enrichment of the halo has been inhomogeneous on timescales $\geq 10^7$
yr (i.e. the lifetimes of $\sim 12$~$M_\odot$ stars), the material
ejected by the more massive stars (on shorter timescales) is considered
well mixed. Consequently, to take a zero Fe yield from high mass SNe
will not affect considerably the results for the inhomogeneous
composition of the gas at t $\geq 10^8$ yr, when Eu production starts.

At the last timesteps, when [Fe/H] is converging to $\simeq -1.5$, the
minimum value of [Eu/Fe] predicted by our model is too low compared
with observations.  Two points must be stressed: first, at [Fe/H] $>
-2$ the chemical enrichment of the gas starts to be dominated by the
disk, while in our simulation we followed the evolution of the halo gas
only. Second, at these epochs we need to take into account other
stellar sources for the chemical enrichment of the gas, i.e. Type Ia
SNe for Fe, intermediate- and low-mass Asymptotic Red Giant stars for
heavy elements.  To obtain an early Eu enrichment (see Fig.~2, upper
panel) we need to reduce the time delay for the production of Eu. For
this reason we explored different SNe mass ranges, adopting the
corresponding Eu yields computed by Travaglio et al.~(1999).  For
example, if we assume the case of Eu production from high-mass SNe
(15--25~$M_\odot$), the time delay in the enrichment of Eu with respect
to Fe will be too small to match the observed spread in [Eu/Fe] at
$-3.5 \leq$ [Fe/H] $\leq -2.5$.

Finally, we tested the sensitivity of our results to another parameter
of our model, the time interval between two subsequent bursts of star
formation inside clouds. The results presented here were all obtained
with $t_{\rm burst}=2\times 10^6$~yr.  For a longer time interval
between bursts ($t_{\rm burst}\simeq 8\times 10^6$~yr) the enrichment
of Eu in the ISM is slower than the standard case, and the discrepancy
between model results and observational data becomes particularly
evident at the lowest metallicities ([Fe/H] $< -3.5$).  A significative
constraint on $t_{\rm burst}$ can be obtained by applying the same
analysis presented here for Eu to elements observed at even lower
metallicities ([Fe/H] $\simeq -4$), like Ba and Sr. These results will
be presented in a forthcoming paper, together with a more detailed
study of the sites for the production of $r$-process and the analysis
of the consequences of the infall of gas from the halo to the disk.

\beginrefer
\refer Burbidge, E.M., Burbidge, G.R., Fowler, W.A., Hoyle, F., 1957, 
Rev. Mod. Phys., 29, 547

\refer Freiburghaus, C., Rembges, J.F., Rauscher, Kolbe, E.T., 
Thielemann, F.K., Kratz, K.L., Pfeiffer, B., Cowan, J.J., 1999, 
ApJ, 516, 381

\refer Gilroy, K.K., Sneden, C., Pilachowski, C.A., Cowan, J.J., 1988,
ApJ, 327, 298

\refer Gratton, R.G., Sneden, C., 1994, 
A\&A, 287, 927

\refer Hoyle, F., 1953, 
ApJ, 118, 513

\refer Ikuta, C., Arimoto, N., 1999, 
PASJ, in press

\refer Ivans, I.I., Sneden, C., Kraft, R.P., Suntzeff, N.B., 
Smith, V.V., Langer, G.E., Fulbright, J.P., 1999, 
astro-ph/9905370

\refer McWilliam, A., Preston, G.W., Sneden, C., Shectman, S., 1995a, 
AJ, 109, 2736 

\refer McWilliam, A., Preston, G.W., Sneden, C., Searle, L., 1995b, 
AJ, 109, 2736 

\refer McWilliam, A., 1998, 
AJ, 115, 1640

\refer McWilliam, A., Searle, L., 1999 
in Galaxy Evolution:
Connecting the Distant Universe with the Local Fossil Record,
ed. M. Spite,
(Dordrecht: Kluwer), in press

\refer Norris, J.E., Ryan, S.G., Beers, T.C., 1997, 
ApJL, 489, L169

\refer Oort, J.H., 1954, 
Bull. Astron. Inst. Netherlands, Vol.12, p.455

\refer Raiteri, C.M., Villata, M., Gallino, R., Busso, M., 
Cravanzola, A., 1999, 
ApJL, 518, L91

\refer Ryan, S.G., Norris, J.E., 1991, 
AJ, 471, 254

\refer Ryan, S.G., Norris, J.E., Beers, T.C., 1996, 
ApJ, 471,254

\refer Ryan, S.G., Norris, J.E., Bessel, M.S., 1991, 
AJ, 102, 303

\refer Shetrone, M.D., 1996, 
AJ, 112, 1517

\refer Sneden, C., Preston, G.W., McWilliam, A., Searle, L., 1994, 
ApJL, 431, L27

\refer Sneden, C., Cowan, J.J., Burris, D.L., Truran, J.W., 1998, 
ApJ, 496,
235

\refer Travaglio, C., Galli, D., Gallino, R., Busso, M., Ferrini, F., 
Straniero, O., 1999, ApJ, 521, 691

\refer Truran, J.W., 1981, 
A\&A, 97, 391

\refer Tsujimoto, T., Shigeyama, T., Yoshii, Y., 1999, 
ApJL, 519, L63

\refer Wheeler, J.C., Cowan, J.J., Hillebrandt, W., 1998, 
ApJL, 493, L101 

\refer Woosley, S.E., Langer, N., Weaver, T.A., 1995, 
ApJ, 448, 315

\refer Woosley, S.E., Weaver, T.A., 1995, 
ApJS, 101, 181

\endrefer           
\end{document}